\documentclass[aps,twocolumn,prl,superscriptaddress,graphics]{revtex4}

\input epsf
\usepackage{graphicx}

\newcommand\be{\begin{equation}}
\newcommand\ee{\end{equation}}
\newcommand\bea{\begin{eqnarray}}
\newcommand\eea{\end{eqnarray}}
\newcommand\vs{\nonumber\\}

\newcommand{\eql}[1]{\label{#1}}
\newcommand{\ec}[1]{Eq.~(\ref{#1})}
\newcommand{\Ec}[1]{(\ref{#1})}
\newcommand{\rf}[1]{\ref{fig:#1}}

\begin{document}
\bibliographystyle{apsrev}


\title{Scale Dependent Spectral Index in Slow Roll Inflation}


\author{Scott Dodelson}
\affiliation{NASA/Fermilab Astrophysics Center
Fermi National Accelerator Laboratory, Batavia, IL~~60510-0500}
\affiliation{Department of Astronomy \& Astrophysics,
The University of Chicago, Chicago, IL~~60637-1433}
\author{Ewan Stewart}
\affiliation{Department of Physics, KAIST, Taejeon 305-701, South Korea}


\date{\today}

\begin{abstract}
Recent observations suggest that the spectral index of the primordial perturbations 
is very close to unity, as expected in models of slow roll inflation.
It is still possible for such models to produce spectra which are scale dependent.
We present a formula for the spectrum produced by an arbitrary inflaton potential
(within the context of slow roll models);
this formula explicitly illustrates and accounts for the possiblity of scale
dependence. A class of examples are studied and comparisons made with the standard
slow roll formula. 

\end{abstract}
\pacs{}

\maketitle


{\parindent 0pt \bf Introduction}
A wide variety of cosmological observations have recently converged on a standard model of cosmology.
Anisotropies in the Cosmic Microwave Background have been measured on scales ranging from the horizon down to several arcminutes.
Inhomogeneities in the universe have been probed by galaxy surveys \cite{surveys} and by observations of the Lyman $\alpha$ 
forest \cite{croft} in the spectra of distant quasars.
The background cosmology has been explored most notably with the aid of Type Ia Supernovae \cite{supernova}.
These observations (and many others) point to a flat universe \cite{flat} with
(i) non-baryonic dark matter; (ii) dark energy; and (iii) primordial adiabatic perturbations with a spectral index very close to unity.
Here we focus on the implications of the last of these, the nature of the primordial perturbations responsible for structure in the universe.

The COBE experiment first placed strong constraints on the slope of the primordial power spectrum by measuring the anisotropies on large scales.
It restricted the spectral index, $n$, to be $1.2\pm 0.3$ \cite{bennett}.
Combining these large angle results with recent measurements of anisotropies on small angular scales 
\cite{dasi,boomerang,maxima,tegmark} leads to even stronger constraints.
For example, combining COBE with DASI \cite{dasi} leads to $n = 1.01_{-0.06}^{+0.08}$. Similar constraints emerge from Boomerang \cite{boomerang} and Maxima \cite{maxima}.
These experiments cover physical scales ranging from  $k\sim 5 \times 10^{-4}$ h Mpc$^{-1}$ down to $k \sim 0.1$h Mpc$^{-1}$.
The Map and Planck satellites will probe this region with even greater sensitivity, reducing the error bars further.
It is even possible to get information about the primordial power spectrum from smaller scales.
The Lyman-$\alpha$ forest for example contains relatively unprocessed information about the spectrum on scales even smaller than $k=1$h Mpc$^{-1}$ \cite{croft}.
The current constraints on the shape of the primordial spectrum, therefore, will only get stronger over the coming decade.

The theory of inflation \cite{guth} has faired well in this latest round of
cosmological discoveries.
Generically, slow roll inflation predicts that the universe is flat, and that the primordial perturbations are Gaussian, adiabatic, and have a nearly scale invariant spectrum.
The degree to which slow roll inflation predicts a scale invariant spectrum depends upon the dynamics of the scalar field(s) controlling inflation.
The simplest possibility is a single `inflaton', slowing rolling down its potential
with its kinetic energy strongly damped by the Hubble expansion.
In the limit in which the rolling is infinitely slow and the damping infinitely strong, the primordial spectrum is a power law, with index $n$ exactly equal to one.
Deviations from $n=1$ are measures of how slowly the field rolled and how strongly its motion was damped during inflation.
Equivalently, different inflationary models predict different values of $n$ or more generally of the shape of the spectrum; measurements of this primordial spectrum enable one to discriminate among different inflationary models.

There is another reason why precise measurements of the primordial spectrum are
important to proponents of inflation.
Even before inflation was proposed, Harrison and Zel'dovich introduced the notion that scale free ($n=1$) adiabatic perturbations represent natural initial conditions. 
A spectrum with $n$ not exactly equal to one, or even more telling, one with deviations from a pure power law form, is perfectly compatible with inflation.
While not a {\it proof\/} of inflation, such deviations would surely be a {\it disproof\/} of the Harrison-Zel'dovich speculations.

In the slow roll approximation, $|n-1|$ is small.
It is often assumed \cite{turner} that deviations from a pure power law are of order $(n-1)^2$.
If true, this would mean that the recent measurements indicating $|n-1|$ is smaller than about $0.1$ 
imply that deviations from a power law would only show up at the percent level at best.

Here we 
(i) show that power law deviations might be significantly larger than this even within the context of slow roll inflation;
(ii) give explicit formulae for these deviations in terms of the inflaton potential; and (iii) illustrate
the usefulness of these formulae with a class of examples.
These examples serve as a warning against extrapolating current measurements beyond their regime of applicability.
For, if the primordial spectrum is not a pure power law, as we argue it may not be, then we do not have 
much independent information beyond $k \sim 1$h Mpc$^{-1}$.
We conclude by mentioning several ramifications of this ignorance.

{\parindent 0pt \bf Slow Roll Expansion}
During inflation, the inflaton $\phi(t)$, which we assume here is a single real scalar field, 
has time dependence characterized by the dimensionless parameters
\be
\epsilon \equiv \frac{1}{2} \left( \frac{\dot\phi}{H} \right)^2
\qquad \mathrm{and} \qquad
\delta_p \equiv \frac{1}{H^p \dot\phi} \left(\frac{d}{dt}\right)^p \dot\phi
\,,\ee
where a dot denotes the derivative with respect to time, $H$ is the Hubble rate, and we have set $8\pi G=1$.
$\delta_1$, which measures 
the second derivative of $\phi$, is sometimes simply called $\delta$.
The evolution of these parameters is governed by
\be\eql{edot}
\frac{d\epsilon}{d\ln a} = 2(\epsilon+\delta_1)\epsilon
\ee
and
\be\eql{ddot}
\frac{d\delta_p}{d\ln a} = \delta_{p+1} + (p\epsilon-\delta_1)\delta_p
\,.\ee

The slow roll approximation assumes, for some small parameter $\xi$, which
observations suggest is of order $0.1$ or smaller,
\be
\epsilon = \mathcal{O}(\xi) \qquad {\rm and} \qquad \delta_1 = \mathcal{O}(\xi).
\ee
The first of these implies that the energy density $\rho = 3H^2 = V + \dot\phi^2/2 \simeq V$, 
and the second 
that the equation of motion $\ddot\phi + 3H\dot\phi + V' = 0$ reduces to the slow roll 
equation of motion $3H\dot\phi + V' \simeq 0$.
Using Eqs.~(\ref{edot}) and~(\ref{ddot}), for this to remain true over a number of $e$-folds we also require
\be
\delta_p = \mathcal{O}(\xi)
\ee
for $p > 1$.
In this approximation, the spectral index is
\be\eql{nslowroll}
n - 1 = - 4\epsilon - 2\delta_1 - 2 \sum_{p=1}^\infty d_p \delta_{p+1} + \mathcal{O}(\xi^2)
\,,\ee
where the $d_p$ are numerical coefficients of order unity.
The first two terms on the right represent the textbook (e.g. Ref.~\cite{liddle}) result. The class of terms in $\mathcal{O}(\xi^2)$ includes the terms $\epsilon^2$, $\epsilon\delta_1$ and $\delta_1^2$ with some numerical coefficients.
The sum includes the higher derivatives of $\phi$; it is these we will be most concerned with here.
The most important point about \ec{nslowroll} is that it 
shows that the recent determinations that $n$ is close to one 
verify slow roll.
That is, at least in the absence of surprising cancellations (which will not concern us), all the terms on the right must be small since the left hand side has been measured to be small. 

Given the validity of the slow roll approximation, an important question remains about the terms in the sum in \ec{nslowroll}.
Is $\delta_p = \mathcal{O}(\xi^p)$ or is $\delta_p = \mathcal{O}(\xi)$?
Either condition would still satisfy slow roll, so there is as yet no experimental way to favor one over the other.
The former is often assumed.
If this assumption is incorrect, then an analyst using it will map an observation (of $n$) onto the wrong set of parameters $\epsilon,\delta_1$.
More importantly, the deviation of the primordial power spectrum from a power law is often 
described by the {\it running\/} of the spectral index.
This running is equal to (again in the slow roll approximation)
\be
\frac{d n}{d\ln k} = - 2 \sum_{p=0}^\infty d_p \delta_{p+2} + \mathcal{O}(\xi^2)
\,.\ee
The parameters $\epsilon$ and $\delta_1$ appear only quadratically in the running, represented by the $\mathcal{O}(\xi^2)$ on the right.
Therefore, if $\delta_p = \mathcal{O}(\xi^p)$, the running will also be of order $\xi^2$, 
on the border of detectability~\cite{turner}.
On the other hand, models in which $\delta_p = \mathcal{O}(\xi)$ still satisfy slow roll, but produce significant running.
Indeed, in these models, the running is expected to be of order $\xi$, \mbox{i.e.} as large as the deviation of $n$ from one.

{\parindent0pt \bf Slow Roll Results}
It behooves us therefore to determine the spectral index and its running in the general case in which $\delta_p = \mathcal{O}(\xi)$.
Here we simply present the results; a companion paper \cite{stewart} gives derivations.
There it is shown that the $d_p$ are best determined via a generating function.
Explicitly, 
\be
\sum_{p=0}^\infty d_p x^p
=
2^x \cos\left(\frac{\pi x}{2}\right) \frac{\Gamma(2-x)}{1+x}
\,.\ee
This relation uniquely determines the coefficients $d_p$. Some explicit values are
\bea
d_0 &=& 1
\,,\qquad
d_1 = -\alpha
\,,\qquad
d_2 = \frac{\alpha^2}{2} - \frac{\pi^2}{24}
\,,\vs
d_3 &=& -\frac{\alpha^3}{6} + \frac{\alpha\pi^2}{24} - \frac{2}{3}
+ \frac{\zeta(3)}{3}
\,,\eea
where $\alpha\equiv 2-\ln2-\gamma\simeq 0.730,$ $\gamma$ is the Euler-Mascheroni
constant, and $\zeta$ is the Riemann zeta function.

Perhaps even more important for the purposes of testing inflationary models
are expressions for the spectral index and its running in terms of the 
inflaton potential, $V(\phi)$.
Ref.~\cite{stewart} shows that
\bea
P &=& \frac{V^3}{12\pi^2 (V^{(1)})^2} \left\{
1 + \left( 3q_1 - \frac{7}{6} \right) \left( \frac{V^{(1)}}{V} \right)^2
\right. \vs && \left. \mbox{}
- 2 \sum_{p=1}^\infty q_p \left( \frac{V^{(1)}}{V} \right)^{p-1}
\frac{V^{(p+1)}}{V} + \mathcal{O}\left(\xi^2\right)
\right\}
\,.\eql{final}\eea
Here, $V^{(p)}$ denotes the $p^\mathrm{th}$ derivative of $V$ with respect to $\phi$.
The potential and its derivatives in \ec{final} are to be evaluated at the value
$\phi$ had at the time when the mode $k$ left the horizon during inflation,
to be precise at the time when $aH = k$;
that is, different scales $k$ correspond to different values of $\phi$.
The coefficients $q_p$ are again best determined via a generating function.
In this case
\be
\sum_{p=0}^\infty q_p x^p = Q(x) \equiv
2^{-x} \cos\left(\frac{\pi x}{2}\right) \frac{3\Gamma(2+x)}{(1-x)(3-x)}
\,.\ee
Some explicit values are
\bea
q_0 &=& 1
\,,\qquad
q_1 = \alpha + \frac{1}{3}
\,,\qquad
q_2 = \frac{\alpha^2}{2} + \frac{\alpha}{3} - \frac{\pi^2}{24} + \frac{1}{9}
\,,\vs
q_3 &=& \frac{\alpha^3}{6} + \frac{\alpha^2}{6} - \frac{\alpha\pi^2}{24}
+ \frac{\alpha}{9} + \frac{2}{3} - \frac{\zeta(3)}{3} - \frac{\pi^2}{72}
+ \frac{1}{27}
\,.\eea

The spectral index and its running can also be expressed in terms of 
the potential and its derivatives, by differentiating the power spectrum 
with respect to $\ln k$, using $\partial/\partial \ln k \rightarrow
-(V^{(1)}/V)\,\partial/\partial\phi$. 
The spectral index is
\be\eql{npot}
n - 1 = -3 \left(\frac{V^{(1)}}{V}\right)^2 + 2 \sum_{p=0}^\infty q_p
\left(\frac{V^{(1)}}{V}\right)^p \frac{V^{(p+2)}}{V}
,\ee
with running
\be\eql{runpot}
\frac{d n}{d\ln k} = -2 \sum_{p=0}^\infty q_p
\left(\frac{V^{(1)}}{V}\right)^{p+1} \frac{V^{(p+3)}}{V}
\,.\ee

Setting $q_p=0$ for all $p>0$ corresponds to the standard slow roll result
\cite{sl,turner,liddle}.
The first two correction terms have also been derived previously: 
the values of $d_1$ and $q_1$ agree with the results of Ref.~\cite{sl}, 
while $d_2$ and $q_2$ agree with Ref.~\cite{sgong}.

These results can also be expressed as \cite{stewart}
\bea\label{Palt}
P & = & \frac{V^3}{12\pi^2\left(V^{(1)}\right)^2} \left\{ 1
+ \left(3\alpha-\frac{1}{6}\right) \left(\frac{V^{(1)}}{V}\right)^2
\right. \\ && \left. \mbox{}
- 2 \int_0^\infty \frac{du}{u} \left[W(u)-\theta(1-u)\right]
\left.\frac{V^{(2)}}{V}\right|_{aH=k/u} \right\} \nonumber
\,,\eea
where $\theta(x)=0$ for $x<0$ and $\theta(x)=1$ for $x>0$,
\be
n - 1 = - 3 \left(\frac{V^{(1)}}{V}\right)^2
- 2 \int_0^\infty du\,W'(u) \left.\frac{V^{(2)}}{V}\right|_{aH=k/u}
\ee
and
\be
\frac{dn}{d\ln k} = 2 \int_0^\infty du\,W'(u)
\left.\frac{V^{(1)}}{V}\frac{V^{(3)}}{V}\right|_{aH=k/u}
\,,\ee
where $W(x) \equiv (6/x)\,j_1(2x)-3\,j_0(2x)$.
Here, standard slow roll would correspond to setting $V^{(2)}/V$ and
$V^{(1)}V^{(3)}/V^2$ to constants.

{\parindent0pt \bf Examples}
For the purposes of illustration, we now introduce a class of 
models in which the spectral index is {\it not\/} a constant. 
Consider the potential 
\be\eql{pot}
V = V_0 e^{\lambda\phi} \left[ 1 + A f(\nu\phi) \right]
\,\ee
where $\lambda$ and $A$ are small, $\nu$ is large, and $f$ is a smooth function.
A surprisingly large number of models \cite{inflmod} can be parametrized in this
way.

\begin{figure}
\includegraphics[width=8cm]{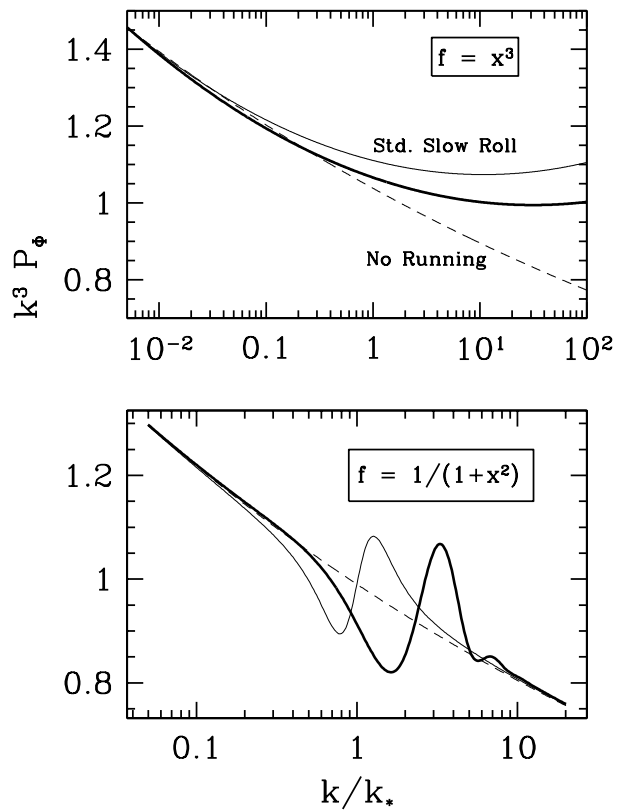}
\caption{Power spectrum of the gravitational potential in two inflationary models
corresponding to potentials of the form in \ec{pot}.
$k_*$ is a fiducial wavenumber depending on the dynamics of the inflaton.
The Harrison-Zel'dovich spectrum is flat.
The thick solid line is the result of \ec{final};
the thin line is the standard slow roll approximation in which $Q\rightarrow 1$;
and the dashed line is the assumption of no running.
The top panel has parameters $\lambda=0.03,\nu=1/\lambda$
while the bottom has $\nu = 7,\lambda=0.3$.
In both cases, $A=\lambda^3/\nu$.}
\label{fig:spec}
\end{figure}

Ref.~\cite{stewart} uses \ec{final} and \ec{Palt} to derive explicit expressions
for the power spectrum and its derivatives when the potential is of the form
\ec{pot}.
Figure~\rf{spec} shows the power spectrum (of the gravitational potential
$\Phi$) in two examples.
In each case, three curves are plotted: the exact result of \ec{final},
the standard slow roll result corresponding to setting $q_p=0$ for all $p>0$,
and the ``no running'' approximation in which the power spectrum is assumed
to be a pure power law. 

There are two important lessons to be learned from Figure~\rf{spec}.
First, and most important, running can be significant, even if deviations from
slow roll -- as determined by measuring the spectral index on large scales --
are small.
When $f(\nu\phi)=\nu^3\phi^3$ (top panel), measuring $n$ on large scales and
extrapolating to small scales with a pure power law underestimates the power
significantly.
We emphasize that
(i) this serious misestimate takes place even though $n$ on large scales is
close to one $(\sim 0.94)$
and
(ii) depending on the parameters in the potential, the estimate could have gone
the other way with a large overestimate of the power.
The second important feature of Figure~\rf{spec} is that the standard slow roll
approximation is not particularly good.
This shows up for the $x^3$ potential, but even more dramatically for the bump
potential in the bottom panel.
Besides the incorrect placement of the bump in the power spectrum and the
too-small amplitude, standard slow roll does not produce any ringing in the
spectrum.
These are already evident for the parameter choice in Figure~\rf{spec} and
become even more pronounced for larger values of $\nu$.
Many groups have studied bumps, dips, and steps in the power spectrum. 
Eqs.~\Ec{final} and~\Ec{Palt} are good ways to analyze these models:
simpler than full numerical solutions and more accurate than standard slow roll.

{\parindent0pt \bf Conclusions}
Inflation, and in particular slow roll inflation, has emerged from the recent confrontation with data in marvelous shape.
Current data support the idea that the universe is flat, and that the primordial power spectrum was close to scale-invariant.
We have shown here that these successes do {\it not\/} necessarily imply that the spectrum is a pure power law on all scales.
Deviations from power law behavior, i.e. running of the spectral index, can be as large as the deviation of the spectral index from unity.
This is exciting, for it suggests that future experiments may be able to measure this running.
If measured, these two sets of small deviations (away from $n=1$ and away from pure power law) will clearly rule out the Harrison-Zel'dovich spectrum.
This would lay waste to one of the most potent arguments today against inflation
(``If the perturbations came from inflation, why did Harrison and Zel'dovich come up with them long before inflation was invented?'').

A related point is that our measurements to date have been predominantly on large scales. It is dangerous to extrapolate these large scale measures of the power to small scales, assuming a pure power law.
Thus, limits on the spectral index from e.g. primordial black holes \cite{blackhole} would be relaxed if the primordial spectrum is not a pure power law. 

More intriguing is the idea that running may help solve some of the small scale problems currently facing Cold Dark Matter \cite{moore}.
It has been suggested that these problems could be alleviated by reducing the small scale power.
Running of the spectral index provides a clean way of doing this.

\begin{acknowledgments}
This work was supported by the DOE, by NASA grant NAG 5-10842 at Fermilab,
by NSF Grant PHY-0079251 at Chicago,
by Brain Korea 21 and KRF grant 2000-015-DP0080.
EDS thanks the SF01 Cosmology Summer Workshop and the Fermilab Theoretical
Astrophysics Group for hospitality.
\end{acknowledgments}


\newcommand\spr[3]{ {\it Phys. Rep.} {\bf #1}, #2 (#3)}
\newcommand\sapj[3]{ {\it Astrophys. J.} {\bf #1}, #2 (#3)}
\newcommand\saj[3]{ {\it Astron. J.} {\bf #1}, #2 (#3)}
\newcommand\sprd[3]{ {\it Phys. Rev. D} {\bf #1}, #2 (#3)}
\newcommand\sprl[3]{ {\it Phys. Rev. Lett.} {\bf #1}, #2 (#3)}
\newcommand\np[3]{ {\it Nucl.~Phys. B} {\bf #1}, #2 (#3)}
\newcommand\snat[3]{ {\it Nature} {\bf #1}, #2 (#3)}
\newcommand\splb[3]{{\it Phys. Lett. B} {\bf #1}, #2 (#3)}
\newcommand\astroph[1]{{\tt astro-ph/#1}}

\end{document}